\documentclass[prl,twocolumn,showpacs,amsmath,amssymb]{revtex4}
\usepackage{dcolumn}
\usepackage{bm}

\usepackage[pdftex]{graphicx,color}

\newcommand\imag{\ensuremath{\mathrm{i}}}

\begin{document}


\title{
Theoretical Description of Coherent Doublon Creation via Lattice Modulation Spectroscopy
}

\date{\today}
\author{Andreas Dirks}
\email{andreas@physics.georgetown.edu}
\affiliation{Department of Physics, Georgetown University, Washington, DC 20057, USA}
\author{Karlis Mikelsons}
\affiliation{Department of Physics, Georgetown University, Washington, DC 20057, USA}
\author{H. R. Krishnamurthy}
\affiliation{Centre for Condensed Matter Theory, Department of Physics,\\
Indian Institute of Science, Bangalore 560012, India, and \\
Jawaharlal Nehru Centre for Advanced Scientific Research, Bangalore 560064, India
}
\author{J. K. Freericks}
\affiliation{Department of Physics, Georgetown University, Washington, DC 20057, USA}
\affiliation{Kavli Institute for Theoretical Physics, Santa Barbara, CA 93106, USA}

\begin{abstract}
Using a recently developed strong-coupling method, we present a comprehensive theory for doublon production 
processes in modulation spectroscopy of a three-dimensional system of ultracold fermionic atoms in an optical lattice
with a trap.
The theoretical predictions compare well to the experimental time traces of doublon production.
For experimentally feasible conditions, we provide a quantitative prediction for the presence of 
a nonlinear "two-photon" excitation at strong modulation amplitudes.
\end{abstract}

\maketitle

\emph{Introduction.} Lattice modulation spectroscopy has advanced to a standard technique in the physics of ultracold atoms in an optical
lattice \cite{Stoeferle2004, Kollath2006, Joerdens2008,
Huber2009,Sensarma2009,Esslinger2010,Hassler2010,Eckstein2010, Korolyuk2010,
Strohmaier2010, Sensarma2010, Xu2011, Greif2011, Taie2012, Tokuno2012,
Tokuno2012R,Lacki2013}. 
Of particular interest is the possibility to measure the Mott gap and the creation and analysis of long-lived doublons in a 
fermionic Hubbard model with strong repulsion \cite{Kollath2006,Joerdens2008,
Huber2009,Sensarma2009, Esslinger2010, Hassler2010, Eckstein2010,
Korolyuk2010, Xu2011, Tokuno2012, Tokuno2012R, Strohmaier2010, Sensarma2010,
Greif2011, Taie2012}.

On the theoretical side, 1D systems have been numerically studied with DMRG
\cite{Kollath2006}. Based on Fermi's golden rule, it has also been possible
to utilize an equilibrium theory to estimate correct doublon production rates
\cite{Tokuno2012, Tokuno2012R}. Nonequilibrium dynamical mean-field theory calculations
have analyzed models with features analogous to lattice modulation spectroscopy by including time-dependent hopping and 
time-dependent interactions and showing how they affect the double occupancy \cite{Eckstein2010}.
In the linear-response limit, quantum Monte-Carlo calculations have studied phase correlations
between the double occupancy and the lattice modulation \cite{Xu2011}.
However, two- and three-dimensional out-of-equilibrium computations remain a challenging problem for
experimental systems in a trap.

In a Mott insulator, doubly occupied sites can be interpreted as occupied by "doublon" quasi-particles. Their long life time is due to 
a separation of energy scales which requires a rather rare high-order (in the hopping) many-body process for a decay to 
occur \cite{Sensarma2010}.
In this paper, we provide a computational study of the creation of doublons
due to lattice depth modulation.
We derive time-dependent tight-binding parameters for a modulated lattice and
then apply a recently developed computational 
strong-coupling method \cite{Mikelsons2012}.
We then validate our approach by making contact with experimental data by Greif et al.~\cite{Greif2011} who have provided a detailed measurement of the
time evolution of the creation process of doublons in a $^{40}$K system. 
The trap is treated within the local-density approximation (LDA).
Exploring the parameter space further, we find that for sufficiently high modulation amplitudes, processes involving the nonlinear combination of two 
coherent quanta of the many-body system enhanced doublon production rates at a frequency which equals precisely half the value of 
the Hubbard repulsion.
Higher order nonlinear effects are difficult to produce due to the way the amplitude modulation of the potential translates into the time dependence 
of the microscopic parameters of the single-band Hubbard model.

\emph{Method.}
The time dependence of the lattice depth modulation is set as follows:
\begin{equation}
V(t) = V_0 + \chi_{[0,t_\text{mod}]}(t)  \Delta V \sin \omega t,
\label{eq:Vt}
\end{equation}
where $V_0$, $\Delta V,$ and $\omega$ are the average value, modulation amplitude, and modulation frequency of the optical lattice
potential depth, respectively, and 
\begin{equation}
\chi_{[0,t_\text{mod}]}(t) =
\begin{cases}1, &\text{if } t \in [0,t_\text{mod}], \\ 0, &\text{otherwise}\end{cases}
\end{equation}
is the characteristic function of the time interval over which
the lattice depth is modulated. The modulation time period length $t_\text{mod} = n_\text{mod}\cdot 2\pi/\omega$
is a function of the number of modulation cycles $n_\text{mod}$ chosen for the driving of the system.

The Hamiltonian for a single atom in a $d$-dimensional optical lattice is given by \cite{blochreview}
\begin{equation}
H_\text{single}(t) = -\frac{\hbar^2}{2m} \vec \nabla^2 + V(t) \cdot \sum_{i=1}^d \sin^2(kx_i),
\end{equation}
where $k=2\pi/\lambda$ is the lattice vector, with the laser wavelength $\lambda$. 
A natural energy unit to use is the recoil-energy $E_R=\hbar^2k^2/2m$.
Using the respective time-dependent maximally localized Wannier functions 
\cite{Kohn1959}, we map the Hamiltonian to a single-band lattice model. Note that for certain frequencies and 
amplitudes, transitions to higher bands will eventually become important. For inter-band transitions
in particular, corrections from terms coming from the time-derivative of the Wannier functions have to be taken into
account \cite{Lacki2013, Sowinski2012}. But this is not needed for the case we evaluate here, as we always keep the system in the single-band limit.

The many-body physics of fermionic atoms with spin 1/2 is then described 
by the single-band fermionic Hubbard model  \cite{Hubbard1963}
\begin{equation}
\begin{split}
H(t) =&\, -J(t) \sum_{\langle
i,j\rangle,\sigma}\left(c^\dagger_{i\sigma}c_{j\sigma} +\text{h.c.} \right) 
\\&+ U(t) \sum_i n_{i\uparrow} n_{i\downarrow}+\sum_{i\sigma} \epsilon_i(t) n_{i\sigma} ,
\end{split}
\end{equation}
where $J(t)= -\langle
w_i(t)|H_\text{single}(t)|w_j(t)\rangle$ is the hopping between Wannier states $w_i(t)$ and $w_j(t)$
at neighboring sites $i$ and $j$, $\epsilon_i(t)=\tilde \epsilon(t)-U(t)/2-\mu + \omega_\text{trap}^2 r_i^2/2m$ is
the on-site energy with $\tilde \epsilon(t) = \langle
w_i(t)|H_\text{single}(t)|w_i(t)\rangle$, and $U(t)=g \int \mathrm{d}^dr
|w_i(\vec r;t)|^4$ is the time-dependent repulsion of atoms, while $g=4\pi
\hbar^2 a/m$ is determined by the $s$-wave scattering length $a$ 
\cite{Esslinger2010}. 
The bracket $\langle \cdot, \cdot \rangle$ denotes nearest-neighbor pairs on the lattice.
In the experimental comparison, we describe the effects of the trap potential $V_\text{trap}$ within the LDA
by averaging about the respective chemical potentials.
\footnote{
Note that the assumed time-dependence of the single-particle energies is unimportant,
since in this work, we perform calculations on a homogeneous, translationally invariant lattice with a fixed density of fermions,
and in this case equal-time expectation values are independent of the time evolution of $\epsilon(t)$. 
}

In order to compute non-equilibrium observables as a function of time we use a strong-coupling approach
which self-consistently expands the self-energy to second order in the hopping \cite{Mikelsons2012}.
The formalism enables us to numerically evaluate the on-site contour-ordered Green's function 
\begin{equation}
G_\sigma(t,t') = -\imag \left\langle
T_\mathcal{C} c_\sigma(t) c_\sigma^\dagger(t') \right\rangle, 
\end{equation}
where $t$ and $t'$ are times located on the Kadanoff-Baym-Keldysh contour $\mathcal{C}$ \cite{Mikelsons2012}.
In order to evaluate the site's double occupancy $D(t) = \langle n_{\uparrow}n_{\downarrow}\rangle(t)$, we use the following relation for 
its equal-time derivative, where $t<_\mathcal{C}t'$:
\begin{equation}
\left.\frac{\partial G_\sigma(t,t')}{\partial t}\right|_{t'=t^+} = U(t) D(t) + \epsilon(t) \langle n_\sigma\rangle(t) + e^\text{kin}_\sigma(t).
\end{equation}
In this expression, the contribution of the spin state $\sigma$ to the kinetic energy per atom can be
evaluated in momentum space via 
\begin{equation}
e^\text{kin}_\sigma(t) = -2\frac{J(t)}{N_k}\sum_k^{N_k}\langle n_{k\sigma}\rangle (t) \sum_m^d \cos k_m
. 
\end{equation}

\emph{Comparison to Experiment.}
\begin{figure}
\centering
\includegraphics[width=0.94\linewidth]{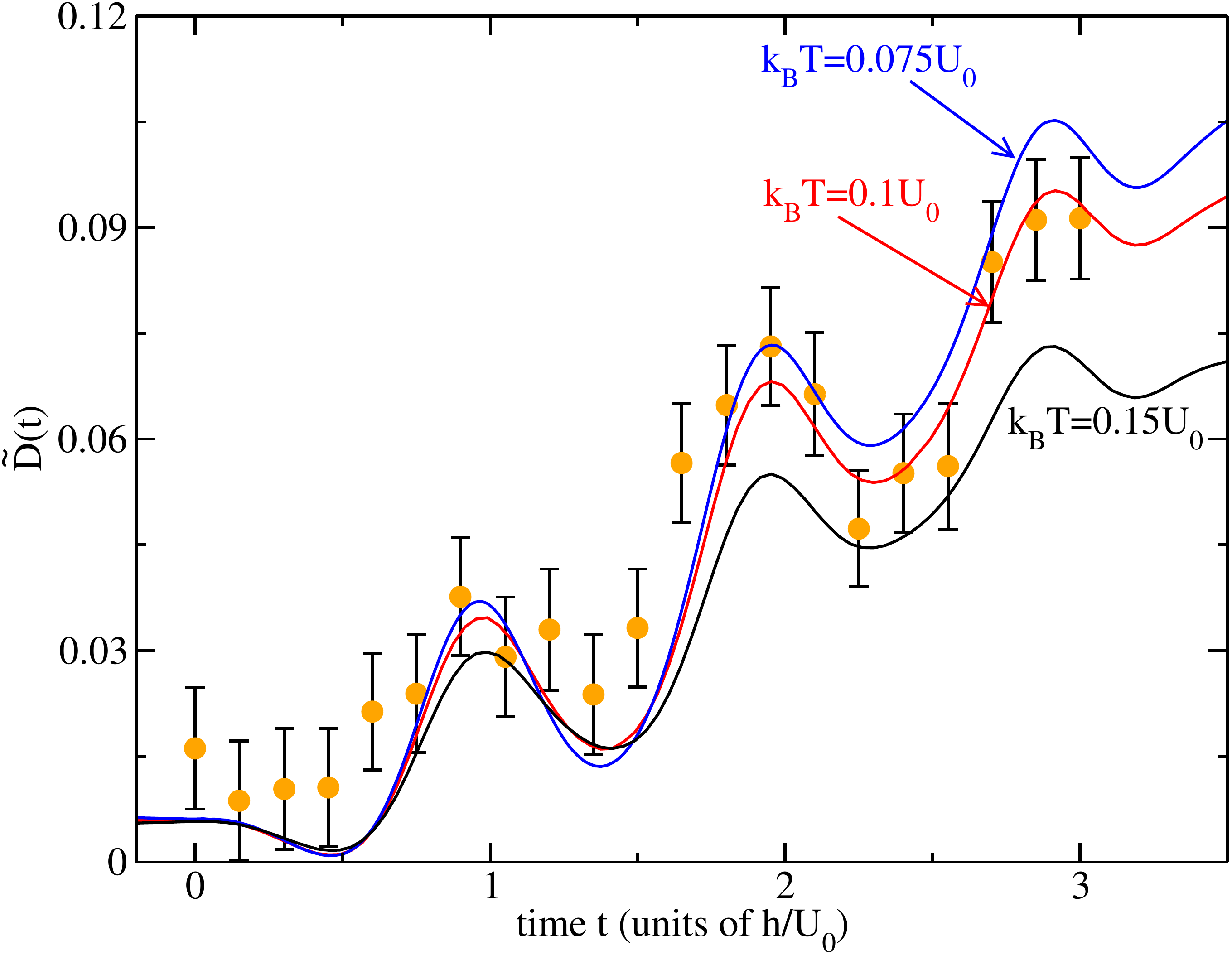}
\caption{(color online) Comparison of the fraction of atoms on doubly occupied sites $\tilde D(t)$ as a function of time to the experimental data from Ref.~\cite{Greif2011}.
Given parameters of the experiment are $U_0/6J_0=4.1$, $V_0=7E_R$, $\Delta V/V_0=20\%$, $\hbar\omega=U_0$. The number of atoms is $80(7)\cdot 10^3$ and the
trap frequency is $\omega_\text{trap}=70.1(5)\cdot2\pi\,s^{-1}$.
Since the initial temperature $T$ is the only unknown parameter, we plot results of our theory for several values of $T$.
The temperature increases from top to bottom as indicated by the labels.
The measurement data were corrected by an offset of $0.0075$ which is due to imperfections
in the preparation of the initial spin mixture \cite{Joerdens2008, pconv}.
The error bars include statistical errors from multiple measurements, as well as an uncertainty of $0.0025$ in the experimental 
determination of the systematic offset. 
}
\label{fig:expcompa}
\end{figure}
In order to validate the strong-coupling approach, Fig.~\ref{fig:expcompa} provides a comparison to recent  
experimental data on the process of doublon creation due to lattice
modulation in a 3D Hubbard model with $^{40}$K \cite{Greif2011}. The figure shows the fraction of atoms on doubly occupied sites
$\tilde D(t) = 2 \sum_i \langle n_\uparrow n_\downarrow \rangle / N$
as a function of time as the lattice depth $V(t)$ is modulated, where $N$ is the number of atoms.
The initial value of $V(t)$ is $V_0=7.0(7)E_R$, and it is modulated by $\Delta V/V_0 = 20\%$ at a driving frequency $\hbar\omega=U_0$. The two-body scattering 
length is tuned through a Feshbach resonance such that $U_0/6J_0=4.1$.
We employ the LDA to use runs at 40 different chemical potentials between $-2U_0$ and $2U_0$, for the experiment's trap
frequency $\frac{\omega_\text{trap}}{2\pi}=70.1\,$Hz and $8\cdot10^{4}$ atoms. We have shown the LDA to be 
quite accurate for a similar 2D system \cite{PRE}.
The temperature is the only unknown parameter in the experiment. For this reason, we plot the LDA results for a set of initial temperatures.
The experimental data have been corrected by a systematic offset between $+0.005$ and $+0.01$ which is known to occur due to imperfections 
in the preparation of the initial spin mixture \cite{Joerdens2008, pconv}.
A reasonable agreement is obtained for temperatures around $k_BT=0.1U_0$. 
We expect the further experimental uncertainties in the particle number and the lattice depth to only add uncertainty to the doublon generation rate.

Since the lattice modulation 
frequency equals the Hubbard repulsion $U_0=U(0)$, particles are resonantly excited from the lower to upper Hubbard 
band during the lattice modulation. There also exists a process of de-excitation which is represented by the decreasing sections of the
observed curve. For the system studied in Fig.~\ref{fig:expcompa}, the excitation process dominates.

\begin{figure}
\centering
\includegraphics[width=0.94\linewidth]{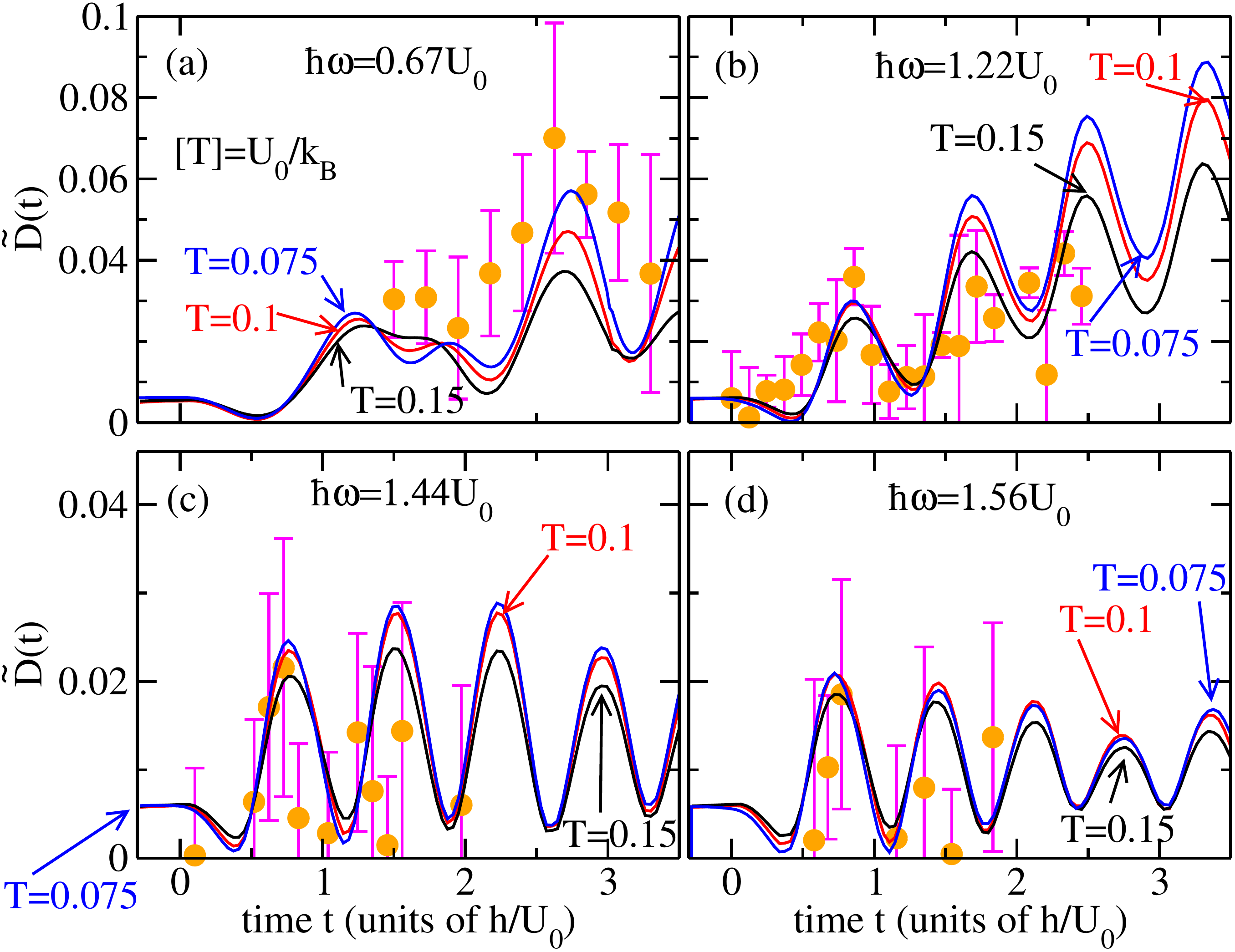}
\caption{(color online) Same as Fig.~\ref{fig:expcompa} for off-resonant modulation frequencies. Data provided by the Esslinger group.
In contrast to Fig.~\ref{fig:expcompa}, the error bars are not statistical errors but spanned by three independent measurements. 
Again, the uncertainty in the systematic offset has been added.
}
\label{fig:freqscan}
\end{figure}

Fig.~\ref{fig:freqscan} shows the same analysis for different modulation frequencies. Clearly, the doublon production rates at these 
frequencies are much lower than for the resonant case $\hbar \omega = U_0$. 
As a consequence, the signal-to-noise ratio in the experiment is also lower.
The temperature dependence of these off-resonant results is weak.

\emph{Amplitude Effects.}
\begin{figure}
\centering
\includegraphics[width=0.94\linewidth]{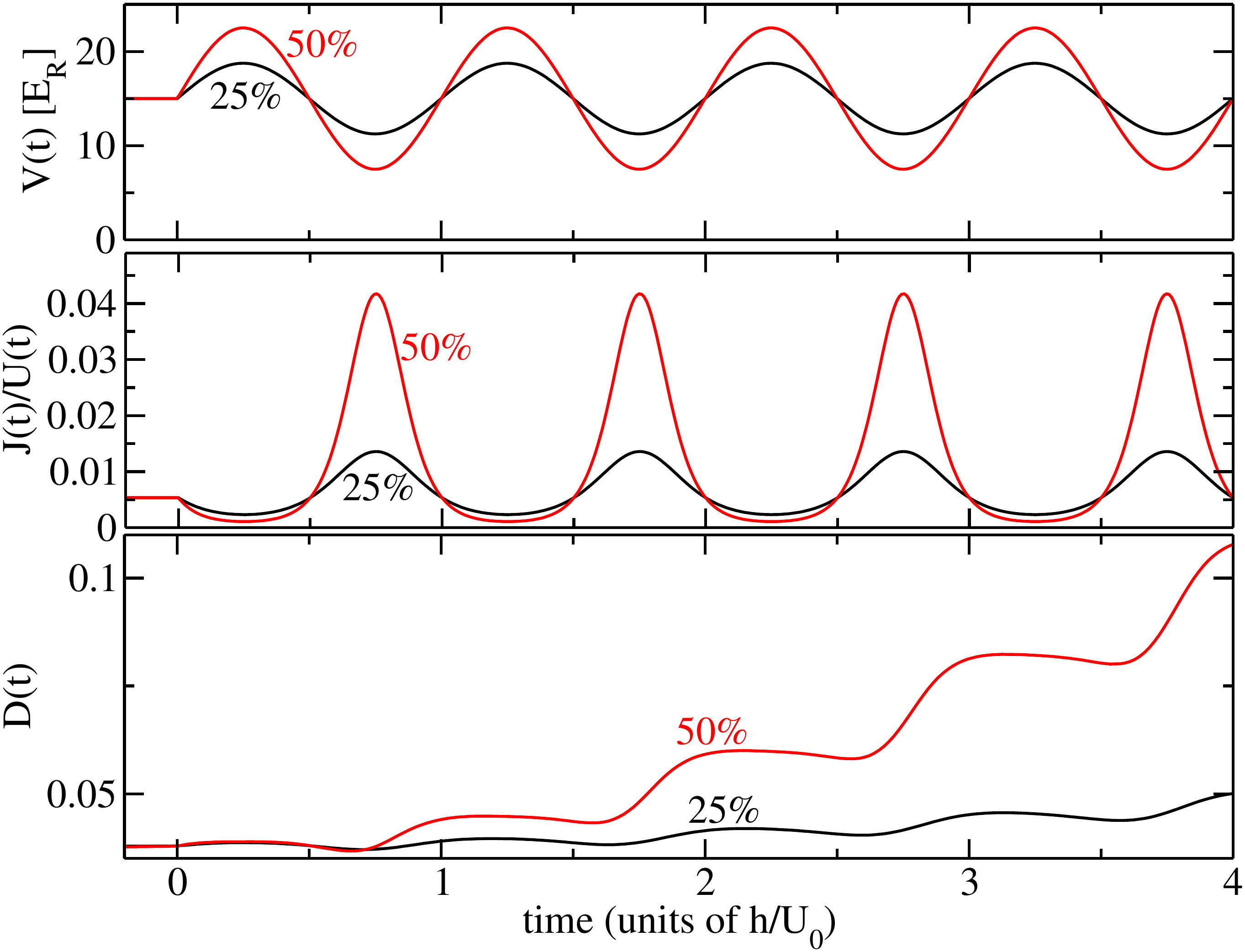}
\caption{(color online) Effect of the amplitude on lattice parameters and doublon production. Model parameters are
$V_0=15E_R$, $U_0/6J_0=31.10$, $k_BT=0.2U_0$, $\hbar\omega=U_0$. The modulation amplitude is either 25\% or 50\%, as specified for each
graph.}
\label{fig:nonlinear}
\end{figure}
Let us now discuss the effect of the amplitude $\Delta V$ in more detail 
for a half-filled system.
In order to prevent inter-band transitions from becoming important, we study this problem for a deeper lattice, such that $\min_{t\in[0,t_\text{mod}]} V(t) \geq 7E_R$.
We have plotted the
lattice depth $V(t)$, the normalized hopping $J(t)/U(t)$, and the double occupancy $D(t)$ as a function of time for 
a deeper lattice at two strong values of the lattice modulation amplitude in Fig.~\ref{fig:nonlinear} for 
a homogeneous system at half filling. 
As the amplitude is raised, the non-linear relationship between the hopping and the lattice depth results
in a periodically kicked rather than a periodically driven system (due to $J(t)/U(t)$ becoming very small for deep lattices).
Overall, the double occupancy is increased stepwise within each modulation cycle.
Again, we observe both, excitation and de-excitation processes in the double occupancy data. Within the first modulation cycle, we observe the
former and the latter to coincide with the decreasing and increasing regimes of $J(t)/U(t)$, respectively.
Within modulation cycles further out in time, the stepwise increase in double occupancy starts already when $J(t)/U(t)$ is still rising.
This is due
to the fact that between the spikes in $J(t)/U(t)$, the hopping is effectively zero, so that the system oscillates internally 
with frequency $U(t)\approx U_0 = \hbar\omega$ within the four-level Hilbert space associated with a single lattice site
\footnote{In this nonlinear regime, the average of $U(t)$ becomes slightly smaller
than $U_0$. The variation in $U(t)$ is much smaller than that of $J(t)$, when the amplitude of the modulation is $50\%$.
For example, when in units of $J_0$, $J(t)$ varies by more than a $100\%$, since it is nearly completely suppressed when $V(t)$ is large and increased by up to $400\%$ when $V(t)$ is small, while $U(t)$ varies 
by only $30\%$ of $U_0$.}.
As a consequence, there is a constructive interference of the immediate effect of lattice modulation and the internal oscillation, 
once the latter is fully engaged.

\emph{Amplitude versus Frequency.}
We would like to address the interplay of the internal oscillation at zero hopping and the lattice modulation in more detail next.
Let us discuss data for a homogeneous half-filled system at different frequencies by first assuming a strong lattice modulation amplitude $\Delta V/V_0=50\%$. 
\begin{figure}
\centering
\includegraphics[width=0.94\linewidth]{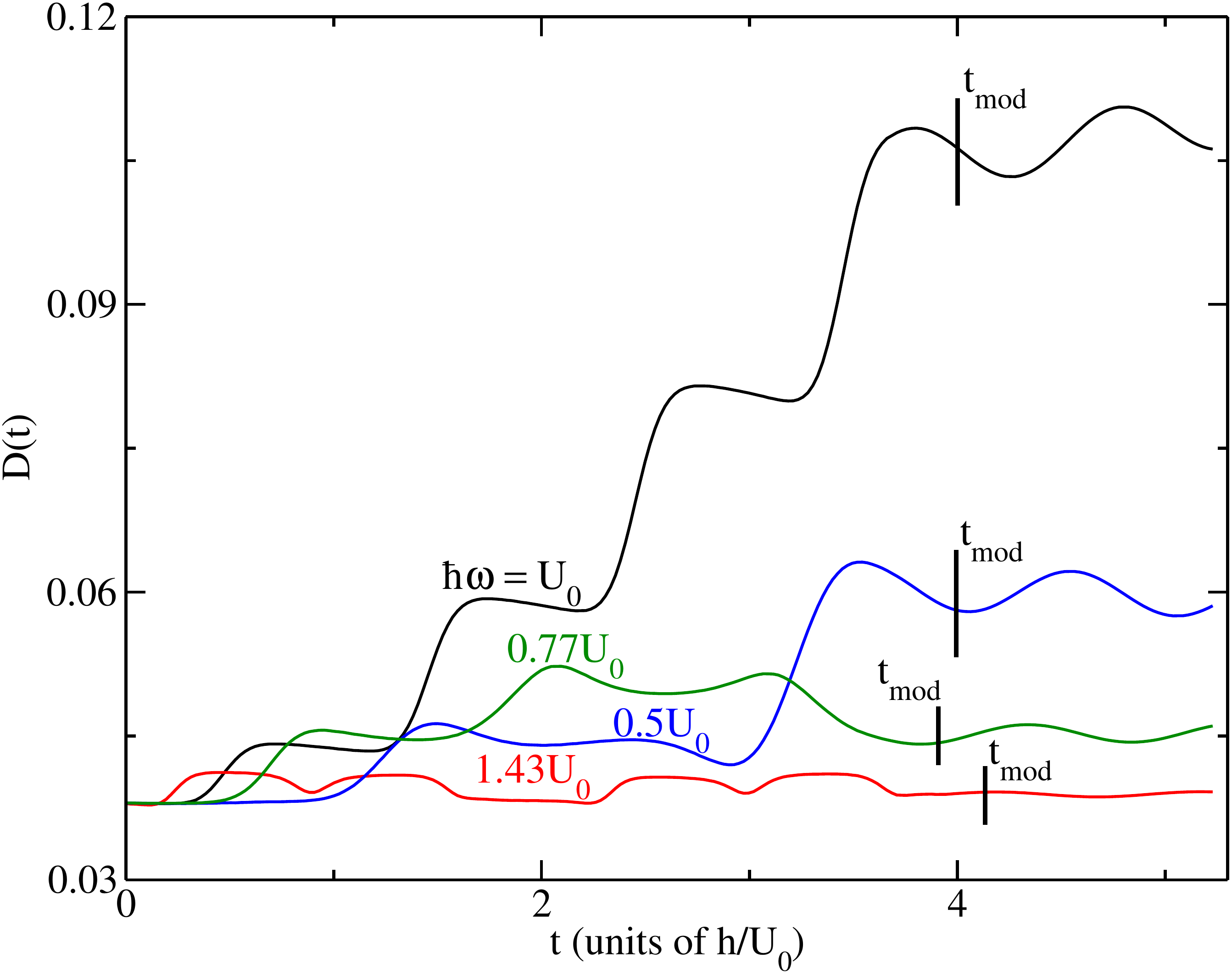}
\caption{(color online) Double occupancy as a function of time for lattice modulations at different frequencies. 
The other model parameters
are same as in Fig.~\ref{fig:nonlinear} for $\Delta V/V_0=50\%$.
Vertical bars denote the respective end time of the modulation.
The modulation frequency associated with a given graph is shown with a label on the curve.
}
\label{fig:timedepresonances}
\end{figure}
The modulation time interval in Eq.~\eqref{eq:Vt} for a given frequency is chosen to be $t_\text{mod} := \left\lfloor\frac{4.93h/U_0}{2\pi/\omega}\right\rfloor\cdot\frac{2\pi}{\omega}$.
Results are shown in Fig.~\ref{fig:timedepresonances}.
The strongest increase in double occupancy is observed in the resonant case 
$\hbar\omega=U_0$. As the end time $t_\text{mod}$ of the modulation is exceeded, the system continues oscillating with the internal 
frequency $U_0$. A significant increase in double occupancy is also observed in the case $\hbar \omega=U_0/2$. Moreover, in this
case also the system continues oscillating with $U_0$ after modulation, but with a smaller amplitude. For the off-resonant frequencies
$\hbar\omega=1.43U_0$ and $\hbar\omega=0.77U_0$,  some intermediate excitations to the upper Hubbard band show up but (partially) annihilate 
after a while. The amplitude of internal oscillations after the modulation is turned off appears to grow with the total increase in 
double occupancy reached when the modulation ends.

\begin{figure}
\centering
\includegraphics[width=0.94\linewidth]{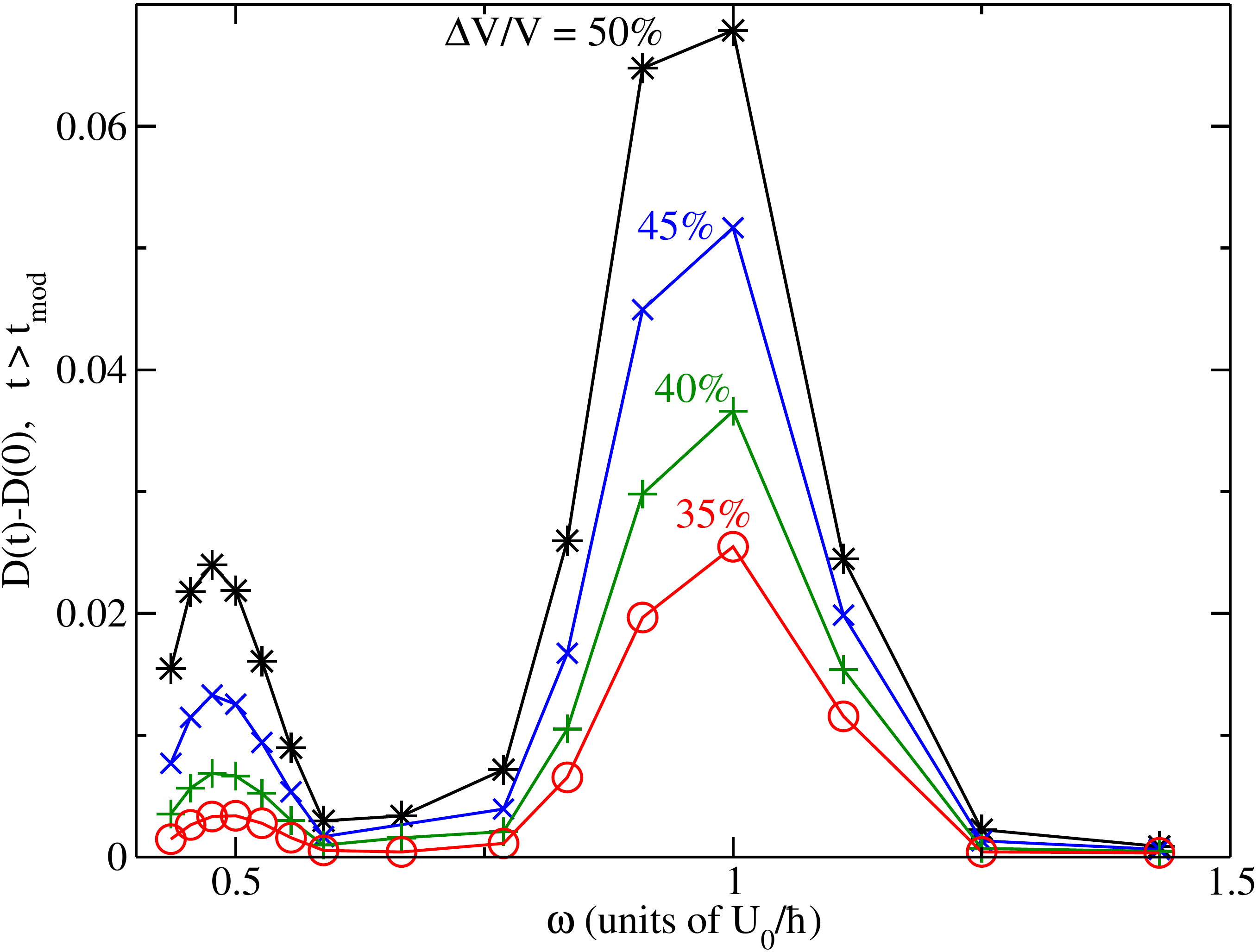}
\caption{(color online) Doublon production as a function of frequency for different amplitudes. Amplitude increases from bottom to top. 
The other parameters are the same as in Figs.~\ref{fig:nonlinear} and  \ref{fig:timedepresonances}.}
\label{fig:freqdep}
\end{figure}

To provide a further overview of the frequency and amplitude dependence, Fig.~\ref{fig:freqdep} shows the final 
value of the double occupancy, i.e. the 
value averaged over one oscillation cycle $[t_\text{mod}, t_\text{mod}+h/U_0]$, vs. the frequency for different amplitudes as a 
doublon production "spectroscopy".
Evidently, a second-order resonance is observed at the frequency $\hbar\omega=0.5U_0$, which is suppressed as the amplitude is lowered.
It can be interpreted as a coherent excitation involving the nonlinear combination of two smaller quanta of energy $U_0/2$, i.e. an
analogue to two-photon excitations in quantum optics. 
A similar effect has already been observed experimentally in a bosonic system.\cite{Mark2011}
As a consequence of the combination 
of these two quanta, the width of the second-order peak is approximately half of the width of the first-order peak.
Also note that both peaks are shifted towards slightly smaller frequencies due to the finite width
of the Hubbard bands. Lower-energy excitations are possible by exciting from the upper edge of the lower 
Hubbard band to the lower edge of upper Hubbard band.

Whereas the $U_0/2$ peak is clearly visible for very large modulation amplitudes, the rather isolated time evolution of lattice sites between 
the kicks in $J(t)/U(t)$ presumably suppresses its amplitude. This is inherent to modulation spectroscopy due to the nonlinearity of the map 
$V \to J/U$. Its presence could be possibly enhanced by designing $V(t)$ in such a way that a harmonic shape is obtained for $J(t)$.

\emph{Summary.}
We have studied finite-amplitude lattice depth modulation spectroscopy of ultracold fermionic atoms in the Mott-insulating phase using a recently 
developed strong-coupling method \cite{Mikelsons2012} in the LDA.
In order to validate the theory, we have compared to experimental data and found good agreement. Only the 
temperature of the initial thermal state was unknown in the experiment and had to be determined \emph{a posteriori}.
We furthermore analyzed higher amplitudes of the modulation strength for deeper lattices.
A large value of the amplitude results in a pulsed, rather than a 
driven system, in terms of the time-dependence of the hopping. This causes step-shaped changes in the double occupancy, accompanied
by oscillations with frequency $U(t)\approx U_0$ of the local degrees of freedom on a single lattice site. At a certain threshold in amplitude, a second 
"nonlinear" peak in the doublon production rate appears at $\omega = U_0/2$.

\section {Acknowledgments}
We thank D. Greif, R. J\"ordens, and T. Esslinger for providing the experimental 
data for Fig.~\ref{fig:freqscan} and for useful discussions.
This work was supported by a
MURI grant from the Air Force Office of Scientific Research
numbered FA9559-09-1-0617. Supercomputing resources
came from a challenge grant of the DoD at the Engineering Research
and Development Center and the Air Force Research and
Development Center. The collaboration was supported
by the Indo-US Science and Technology Forum under
the joint center numbered JC-18-2009 (Ultracold atoms).
JKF also acknowledges the McDevitt bequest at Georgetown.
HRK acknowledges support of the Department of 
Science and Technology in India.
This research was  supported in part by the National Science Foundation under
Grant No. PHY11-25915 during a visit by J.K.F. to KITP, Santa Barbara.

\end{document}